\documentclass[preprintnumbers,amsmath,amssymbm,prd]{revtex4}
\usepackage{epsfig}
\usepackage{graphicx}

\begin{document}
\title{Quasinormal resonances of a massive scalar field in a near-extremal Kerr black
hole spacetime}
\author{Shahar Hod}
\affiliation{The Ruppin Academic Center, Emeq Hefer 40250, Israel}
\affiliation{ }
\affiliation{The Hadassah Institute, Jerusalem 91010, Israel}
\date{\today}

\begin{abstract}
\ \ \ The fundamental resonances of near-extremal Kerr black holes
due to massive scalar perturbations are derived {\it analytically}.
We show that there exists a critical mass parameter, $\mu_c$, below
which increasing the mass $\mu$ of the field increases the
oscillation frequency $\Re(\omega)$ of the resonance. On the other
hand, above the critical field mass increasing the mass $\mu$
increases the damping rate $\Im(\omega)$ of the mode. We confirm our
analytical results by numerical computations.
\end{abstract}
\bigskip
\maketitle

\section{Introduction}

The uniqueness theorems \cite{un1,un2,un3} imply that the metric
outside a newly born black hole should relax into a Kerr-Newman
spacetime, characterized solely by the black-hole mass, charge, and
angular momentum. The relaxation phase in the dynamics of perturbed
black holes is characterized by `quasinormal ringing', damped
oscillations with a discrete spectrum (see e.g. \cite{Nollert1,Ber1}
for detailed reviews). These characteristic oscillations are then
followed by late-time decaying tails \cite{Tails1,Tails2}.

The black hole quasinormal modes (QNMs) correspond to solutions of
the perturbations equations (the Teukolsky master equation
\cite{Teuk}) with the physical boundary conditions of purely
outgoing waves at spatial infinity and purely ingoing waves crossing
the event horizon \cite{Detw}. Such boundary conditions single out a
{\it discrete} set of black-hole resonances $\{\omega_n\}$ (assuming
a time dependence of the form $e^{-i\omega t}$). In analogy with
standard scattering theory, the QNMs can be regarded as the
scattering resonances of the black-hole spacetime. They thus
correspond to poles of the transmission and reflection amplitudes of
a standard scattering problem in a black-hole spacetime.

Quasinormal resonances are expected to be excited by a variety of
astrophysical processes involving black holes. Being the
characteristic 'sound' of the black hole itself, these free
oscillations are of great importance from the theoretical
\cite{HodPRL,Gary} and astrophysical point of view
\cite{Nollert1,Ber1}. They allow a direct way of identifying the
spacetime parameters, especially the mass and angular momentum of
the black hole. This has motivated a flurry of research during the
last four decades aiming to compute the resonance spectrum of
various types of black holes \cite{Nollert1,Ber1}.

It is worth nothing that in most cases of physical interest, the
black-hole QNMs must be computed {\it numerically} by solving the
perturbations equations supplemented by the appropriate physical
boundary conditions. However, it has been shown
\cite{Hod1,Hod2,Hod3} that the spectrum of quasinormal frequencies
can be studied {\it analytically} in the near-extremal limit $a\to
M$, where $M$ and $a$ are the mass and angular momentum per unit
mass of the black hole, respectively.

The dynamics of scalar test fields in black-hole spacetimes is
primarily of theoretical interest-- it usually serves as a toy model
for the analysis of gravitational black-hole perturbations. However,
as pointed out in \cite{Will}, the possible existence of boson stars
could make scalar QNMs observationally relevant. Boson stars are
assumed to be made up of self-gravitating massive scalar fields
\cite{Will,Siedel}. If a boson star becomes unstable and collapses
to form a black hole, it is expected to radiate scalar waves (along
with gravitational waves) in the appropriate QNMs frequencies.

Former numerical investigations of massive QNMs (see e.g.,
\cite{Will,Dolan}) have found that increasing the mass $\mu$ of the
field increases the oscillation frequency $\Re(\omega)$ of the mode.
Below we shall provide an {\it analytical} explanation for this
phenomena. Furthermore, we shall show that there exists a critical
mass parameter, $\mu_c$, above which increasing the mass $\mu$ of
the field actually increases the damping rate $\Im(\omega)$ of the
mode.

\section{Description of the system}

The physical system we consider consists of a massive scalar field
coupled to a rotating Kerr black hole. The dynamics of a scalar
field $\Psi$ of mass $\mu$ in the Kerr spacetime \cite{Kerr} is
governed by the Klein-Gordon equation
\begin{equation}\label{Eq1}
(\nabla^a \nabla_a -\mu^2)\Psi=0\  .
\end{equation}
(It is worth emphasizing that $\mu$ stands for ${\cal M}G/\hbar c$,
where ${\cal M}$ is the mass of the scalar field. We use units in
which $G=c=\hbar=1$.) One may decompose the field as
\begin{equation}\label{Eq2}
\Psi_{lm}(t,r,\theta,\phi)=e^{im\phi}S_{lm}(\theta;a\omega)R_{lm}(r;a\omega)e^{-i\omega
t}\ ,
\end{equation}
where $(t,r,\theta,\phi)$ are the Boyer-Lindquist coordinates
\cite{Kerr}, $\omega$ is the (conserved) frequency of the mode, $l$
is the spheroidal harmonic index, and $m$ is the azimuthal harmonic
index with $-l\leq m\leq l$. (We shall henceforth omit the indices
$l$ and $m$ for brevity.) With the decomposition (\ref{Eq2}), $R$
and $S$ obey radial and angular equations both of confluent Heun
type coupled by a separation constant $K(a\omega)$
\cite{Heun,Flam,Fiz1,Fiz2,Fiz3}. The sign of $\omega_I$ determines
whether the solution is decaying $(\omega_I<0)$ or growing
$(\omega_I>0)$ in time.

The angular functions $S(\theta;a\omega)$ are the spheroidal
harmonics which are solutions of the angular equation
\cite{Teuk,Flam,Abram,Stro}
\begin{equation}\label{Eq3}
{1\over {\sin\theta}}{\partial \over
{\partial\theta}}\Big(\sin\theta {{\partial
S}\over{\partial\theta}}\Big)+\Big[K-a^2(\omega^2-\mu^2)+a^2(\omega^2-\mu^2)\cos^2\theta-{{m^2}\over{\sin^2\theta}}\Big]S=0\
.
\end{equation}
The angular functions are required to be regular at the poles
$\theta=0$ and $\theta=\pi$. These boundary conditions pick out a
discrete set of eigenvalues $\{K_{lm}\}$ labeled by the integers $l$
and $-l\leq m\leq l$. For $a^2(\omega^2-\mu^2)\lesssim m^2$ one can
treat $a^2(\omega^2-\mu^2)\cos^2\theta$ in Eq. (\ref{Eq3}) as a
perturbation term on the generalized Legendre equation and obtain
the perturbation expansion \cite{Abram}
\begin{equation}\label{Eq4}
K_{lm}-a^2(\omega^2-\mu^2)=l(l+1)+\sum_{k=1}^{\infty}c_ka^{2k}(\omega^2-\mu^2)^k\
\end{equation}
for the separation constants $K_{lm}$. The expansion coefficients
$\{c_k\}$ are given in Ref. \cite{Abram}.

The radial Teukolsky equation is given by \cite{Teuk,Hodcen,Stro}
\begin{equation}\label{Eq5}
\Delta{{d} \over{dr}}\Big(\Delta{{dR}\over{dr}}\Big)+\Big[H^2
+\Delta[2ma\omega-K-\mu^2(r^2+a^2)]\Big]R=0\ ,
\end{equation}
where $\Delta\equiv r^2-2Mr+a^2$ and $H\equiv (r^2+a^2)\omega-ma$.
The zeroes of $\Delta$, $r_{\pm}=M\pm (M^2-a^2)^{1/2}$, are the
black hole (event and inner) horizons.

We are interested in solutions of the radial equation (\ref{Eq5})
with the physical boundary conditions of purely outgoing waves at
spatial infinity and purely ingoing waves at the black-hole horizon
(as measured by a comoving observer) \cite{Dolan}. That is,
\begin{equation}\label{Eq6}
R \sim
\begin{cases}
{1\over r}e^{i\sqrt{\omega^2-\mu^2}y} & \text{ as }
r\rightarrow\infty\ \ (y\rightarrow \infty)\ ; \\
e^{-i (\omega-m\Omega)y} & \text{ as } r\rightarrow r_+\ \
(y\rightarrow -\infty)\ ,
\end{cases}
\end{equation}
where the ``tortoise" radial coordinate $y$ is defined by
$dy=[(r^2+a^2)/\Delta]dr$. These boundary conditions single out a
discrete set of resonances $\{\omega_n\}$ which correspond to the
quasinormal resonances of the massive field \cite{Dolan}.
(We note that, in addition to the QNMs resonances, the massive field
is also characterized by a spectrum of bound states
\cite{Dolan,DamDer,HodSO} which tend to zero at spatial infinity.)

\section{The quasinormal resonances}

It is convenient to define new dimensionless variables
\begin{equation}\label{Eq7}
x\equiv {{r-r_+}\over {r_+}}\ \ ;\ \ \tau\equiv{{r_+-r_-}\over
{r_+}}\ \ ;\ \ \varpi\equiv{{\omega-m\Omega}\over{2\pi T_{BH}}}\ \
;\ \ k\equiv 2\omega r_+\  .
\end{equation}
Here $T_{BH}\equiv {{r_+-r_-}\over{4\pi(r^2_++a^2)}}$ and
$\Omega\equiv {{a}\over{r^2_++a^2}}$ are the temperature and angular
velocity of the black hole, respectively. In terms of these
dimensionless variables the radial equation becomes
\begin{equation}\label{Eq8}
x(x+\tau){{d^2R}\over{dx^2}}+(2x+\tau){{dR}\over{dx}}+VR=0\  ,
\end{equation}
where $V\equiv
H^2/r^2_+x(x+\tau)-K_{lm}+2ma\omega-\mu^2[r^2_+(x+1)^2+a^2]$ and
$H=r^2_+\omega x^2+r_+kx+r_+\varpi\tau/2$.

As we shall now show, the spectrum of massive quasinormal resonances
can be studied analytically in the double limit $a\to M$ and
$\omega\to m\Omega$ (see \cite{Teukms} for the massless case). We
first consider the radial equation (\ref{Eq8}) in the far region
$x\gg \text{max}\{\tau,M(m\Omega-\omega)\}$. Then Eq. (\ref{Eq8}) is
well approximated by
\begin{equation}\label{Eq9}
x^2{{d^2R}\over{dx^2}}+2x{{dR}\over{dx}}+V_{\text{far}}R=0\  ,
\end{equation}
where $V_{\text{{far}}}=(\omega^2-\mu^2)r^2_+x^2+2(\omega
k-\mu^2r_+)r_+x+[-K_{lm}+2ma\omega+k^2-\mu^2(r^2_++a^2)]$. A
solution of Eq. (\ref{Eq9}) that satisfies the boundary condition
(\ref{Eq6}) can be expressed in terms of the confluent
hypergeometric functions $M(a,b,z)$ \cite{Morse,Abram,HodSO}
\begin{equation}\label{Eq10}
R=C_1(2i\sqrt{\omega^2-\mu^2}r_+)^{{1\over 2}+i\delta}x^{-{1\over
2}+i\delta}e^{-i\sqrt{\omega^2-\mu^2}r_+x}M({1\over
2}+i\delta+i\kappa,1+2i\delta,2i\sqrt{\omega^2-\mu^2}r_+x)+C_2(\delta\to
-\delta)\  ,
\end{equation}
where $C_1$ and $C_2$ are constants. Here
\begin{equation}\label{Eq11}
\kappa\equiv {{\omega k-\mu^2r_+}\over{\sqrt{\omega^2-\mu^2}}}\ ,
\end{equation}
and
\begin{equation}\label{Eq12}
\delta^2\equiv k^2+2ma\omega-K_{lm}-{1\over 4}-\mu^2(r^2_++a^2)\  .
\end{equation}
The notation $(\delta\to -\delta)$ means ``replace $\delta$ by
$-\delta$ in the preceding term."

We next consider the near horizon region $x\ll 1$. The radial
equation is given by Eq. (\ref{Eq8}) with $V\to
V_{\text{near}}\equiv
-K_{lm}+2ma\omega-\mu^2(r^2_++a^2)+(kx+\varpi\tau/2)^2/x(x+\tau)$.
The physical solution obeying the ingoing boundary conditions at the
horizon is given by \cite{Morse,Abram,HodSO}
\begin{equation}\label{Eq13}
R=x^{-{i\over 2}\varpi}\Big({x\over \tau}+1\Big)^{i({1\over
2}\varpi-k)}{_2F_1}({1\over 2}+i\delta-ik,{1\over
2}-i\delta-ik;1-i\varpi;-x/\tau)\  ,
\end{equation}
where $_2F_1(a,b;c;z)$ is the hypergeometric function.

The solutions (\ref{Eq10}) and (\ref{Eq13}) can be matched in the
overlap region $\text{max}\{\tau,M(m\Omega-\omega)\}\ll x\ll 1$. It
is worth emphasizing that in order to have a non-trivial overlap
region we must restrict our analytical solution to the regime of
rapidly rotating ({\it near-extremal}) black holes. In particular,
the condition $\tau\ll 1$ is satisfied in the near-extremal limit.
The
$x\ll 1$ limit of Eq. (\ref{Eq10}) yields \cite{Morse,Abram,HodSO}
\begin{equation}\label{Eq14}
R\to C_1(2i\sqrt{\omega^2-\mu^2}r_+)^{{1\over 2}+i\delta}x^{-{1\over
2}+i\delta}+C_2(\delta\to -\delta)\  .
\end{equation}
The $x\gg \tau$ limit of Eq. (\ref{Eq13}) yields
\cite{Morse,Abram,HodSO}
\begin{equation}\label{Eq15}
R\to \tau^{{1\over
2}-i\delta-i\varpi/2}{{\Gamma(2i\delta)\Gamma(1-i\varpi)}\over{\Gamma({1\over
2}+i\delta-ik)\Gamma({1\over 2}+i\delta-i\varpi+ik)}}x^{-{1\over
2}+i\delta}+(\delta\to -\delta)\  .
\end{equation}
By matching the two solutions in the overlap region one finds
\begin{equation}\label{Eq16}
C_1=\tau^{{1\over
2}-i\delta-i\varpi/2}{{\Gamma(2i\delta)\Gamma(1-i\varpi)}\over{\Gamma({1\over
2}+i\delta-ik)\Gamma({1\over
2}+i\delta-i\varpi+ik)}}(2i\sqrt{\omega^2-\mu^2}r_+)^{-{1\over
2}-i\delta}\  ,
\end{equation}
\begin{equation}\label{Eq17}
C_2=\tau^{{1\over
2}+i\delta-i\varpi/2}{{\Gamma(-2i\delta)\Gamma(1-i\varpi)}\over{\Gamma({1\over
2}-i\delta-ik)\Gamma({1\over
2}-i\delta-i\varpi+ik)}}(2i\sqrt{\omega^2-\mu^2}r_+)^{-{1\over
2}+i\delta}\  .
\end{equation}

Approximating Eq. (\ref{Eq10}) for $x\to\infty$ one gets
\cite{Morse,Abram,HodSO}
\begin{eqnarray}\label{Eq18}
R&\to&
\Big[C_1(2i\sqrt{\omega^2-\mu^2}r_+)^{i\kappa}{{\Gamma(1+2i\delta)}\over{\Gamma({1\over
2}+i\delta+i\kappa)}}x^{-1+i\kappa}+C_2(\delta\to
-\delta)\Big]e^{i\sqrt{\omega^2-\mu^2}r_+x}\nonumber
\\&& + \Big[C_1(2i\sqrt{\omega^2-\mu^2}r_+)^{-i\kappa}{{\Gamma(1+2i\delta)}\over{\Gamma({1\over
2}+i\delta-i\kappa)}}x^{-1-i\kappa}(-1)^{-{1\over
2}-i\delta-i\kappa}+C_2(\delta\to
-\delta)\Big]e^{-i\sqrt{\omega^2-\mu^2}r_+x}\ .
\end{eqnarray}
A free oscillations of the field (a quasinormal resonance) is
characterized by a purely {\it outgoing} wave at spatial infinity.
Thus, the coefficient of the exponent
$e^{-i\sqrt{\omega^2-\mu^2}r_+x}$ in Eq. (\ref{Eq18}) should vanish,
see Eq. (\ref{Eq6}). Taking cognizance of Eqs.
(\ref{Eq16})-(\ref{Eq18}), one finds the resonance condition for the
quasinormal modes of the massive field:
\begin{equation}\label{Eq19}
{{\Gamma(2i\delta)\Gamma(1+2i\delta)(-2i\tau\sqrt{\omega^2-\mu^2}r_+)^{-i\delta}}\over{\Gamma({1\over
2}+i\delta-i\kappa)\Gamma({1\over 2}+i\delta-ik)\Gamma({1\over
2}+i\delta-i\varpi+ik)}}+{{\Gamma(-2i\delta)\Gamma(1-2i\delta)(-2i\tau\sqrt{\omega^2-\mu^2}r_+)^{i\delta}}\over{\Gamma({1\over
2}-i\delta-i\kappa)\Gamma({1\over 2}-i\delta-ik)\Gamma({1\over
2}-i\delta-i\varpi+ik)}}=0\  .
\end{equation}

The resonance condition (\ref{Eq19}) can be solved analytically in
the regime $\tau\ll 1$ with $\omega\simeq m\Omega$. We first write
it in the form
\begin{equation}\label{Eq20}
{{1}\over{\Gamma({1\over 2}-i\delta-i\varpi+ik)}}={\cal D}\times
(-2i\tau\sqrt{\omega^2-\mu^2}r_+)^{-2i\delta}\ ,
\end{equation}
where ${\cal D}\equiv [\Gamma(2i\delta)]^2\Gamma({1\over
2}-i\delta-i\kappa)\Gamma({1\over
2}-i\delta-ik)/[\Gamma(-2i\delta)]^2\Gamma({1\over
2}+i\delta-i\kappa)\Gamma({1\over 2}+i\delta-ik)\Gamma({1\over
2}+i\delta-i\varpi+ik)$. We note that ${\cal D}$ has a well defined
limit as $a\to M$ and $\omega\to m\Omega$.

In the limit $\omega\to m\Omega$, where $\omega$ is almost purely
real, one finds from Eq. (\ref{Eq12}) that $\delta^2$ is also almost
purely real. If $\delta$ is almost purely real and larger than $\sim
1$, then one has
$(-i)^{-2i\delta}=e^{(-i{{\pi}\over{2}})(-2i\delta)}=e^{-\pi\delta}\ll
1$. If $\delta$ is almost purely imaginary with a positive imaginary
part, then one has $\tau^{-2i\delta}\to 0$ in the near-extremal
$\tau\to 0$ limit. In both cases one therefore finds $\epsilon\equiv
(-2i\tau\sqrt{\omega^2-\mu^2}r_+)^{-2i\delta}\ll 1$ on the r.h.s of
Eq. (\ref{Eq20}).

Thus, a consistent solution of the resonance condition (\ref{Eq20})
may be obtained if $1/\Gamma({1\over
2}-i\delta-i\varpi+ik)=O(\epsilon)$ \cite{Notedelta}. Suppose
\begin{equation}\label{Eq21}
{1\over 2}-i\delta-i\varpi+ik=-n+\eta\epsilon+O(\epsilon^2)\ ,
\end{equation}
where $n\geq 0$ is a non-negative integer and $\eta$ is a constant
to be determined below. Then one has
\begin{equation}\label{Eq22}
\Gamma({1\over
2}-i\delta-i\varpi+ik)\simeq\Gamma(-n+\eta\epsilon)\simeq
(-n)^{-1}\Gamma(-n+1+\eta\epsilon)\simeq\cdots\simeq [(-1)^n
n!]^{-1}\Gamma(\eta\epsilon)\  ,
\end{equation}
where we have used the relation $\Gamma(z+1)=z\Gamma(z)$
\cite{Abram}. Next, using the series expansion
$1/\Gamma(z)=\sum_{k=1}^{\infty} c_k z^k$ with $c_1=1$ [see Eq.
$(6.1.34)$ of \cite{Abram}], one obtains
\begin{equation}\label{Eq23}
1/\Gamma({1\over 2}-i\delta-i\varpi+ik)=(-1)^n
n!\eta\epsilon+O(\epsilon^2)\ .
\end{equation}
Substituting (\ref{Eq23}) into (\ref{Eq20}) one finds $\eta={\cal
D}/[(-1)^n n!]$.

Finally, substituting $\varpi\equiv(\omega-m\Omega)/2\pi T_{BH}$ and
$k\equiv 2\omega r_+=m+O(MT_{BH})$ [the last equality holds for
$\omega=m\Omega+O(T_{BH})$] into Eq. (\ref{Eq21}), one obtains a
simple formula for the quasinormal resonances of the massive field:
\begin{equation}\label{Eq24}
\omega=m\Omega+2\pi T_{BH}[m-\delta-i(n+{1\over
2})]+O(MT^2_{BH},\epsilon T_{BH})\ .
\end{equation}

\section{Numerical confirmation}

We shall now verify the validity of the analytically derived formula
(\ref{Eq24}) for the massive resonances. The black-hole quasinormal
frequencies can be computed using standard numerical techniques, see
\cite{Dolan} for details. We present here results for the case
$l=m=2$. Substituting in Eq. (\ref{Eq4}) $a\to M$ and
$M\omega=1+O(MT_{BH})$ [see Eq. (\ref{Eq24}) for $a\to M$], one
finds $K_{22}\simeq 6{6\over 7}-{6\over 7}M^2\mu^2$, where we have
used the expansion coefficients $c_1={1\over 7}$ and $c_2=-{2\over
1029}$ from \cite{Abram}. Next, substituting this value of $K_{22}$
into (\ref{Eq12}), one obtains
\begin{equation}\label{Eq25}
\delta^2_{22}={{25}\over{28}}-1{1\over 7}M^2\mu^2+O(MT_{BH})\  .
\end{equation}
One therefore finds that $\delta_{22}$ is {\it real} for
$\mu<\mu_c=\sqrt{25/32}M^{-1}$ and {\it imaginary} for larger values
of the field mass. (It is worth emphasizing that in full units $M
\mu$ stands for the dimensionless ratio $GM{\cal M}/\hbar c=M{\cal
M}/ M^2_{\text{Planck}}$.)

Taking cognizance of Eqs. (\ref{Eq24})-(\ref{Eq25}), one finds that
for small mass values ($\mu<\mu_c$, where $\delta_{22}$ is real),
increasing the mass $\mu$ of the field increases the oscillation
frequency $\Re(\omega)$ of the resonance. On the other hand, for
$\mu>\mu_c$ (where $\delta_{22}$ is imaginary) increasing the mass
$\mu$ of the field increases the damping rate $\Im(\omega)$ of the
mode.

In Table \ref{Table1} we present a comparison between the {\it
analytically} derived massive resonances, Eq. (\ref{Eq24}), and the
{\it numerically} computed frequencies \cite{Dolan}. We find an
almost perfect agreement between the two. Table \ref{Table2}
demonstrates the fact that the agreement between the numerical data
and the analytical formula (\ref{Eq24}) is quite good already at
$a/M=0.9$. This is quite surprising since the assumption $\tau\ll 1$
breaks down for this value of the rotation parameter.
\begin{table}[htbp]
\centering
\begin{tabular}{|c|c|c|c|}
\hline
$M\mu$ & $\Re\omega_{\text{ana}}/\Re\omega_{\text{num}}$ & $\Im\omega_{\text{ana}}/\Im\omega_{\text{num}}$ \\
\hline
\ 0.0\ \ & \ 1.003\ \ &\ 0.983\ \\
\ 0.1\ \ & \ 1.003\ \ &\ 0.983\ \\
\ 0.2\ \ & \ 1.003\ \ &\ 0.983\ \\
\ 0.3\ \ & \ 1.004\ \ &\ 0.983\ \\
\hline
\end{tabular}
\caption{Massive scalar quasinormal resonances of a near-extremal
Kerr black hole with $a/M=0.995$. The data shown is for the mode
$l=m=2$, see also \cite{Dolan}. We display the ratio between the
analytically derived frequencies, $\omega_{\text{ana}}$, and the
numerically computed values, $\omega_{\text{num}}$. The numerically
computed frequencies of the massive field agree with the analytical
formula (\ref{Eq24}) to within $2\%$.} \label{Table1}
\end{table}

\begin{table}[htbp]
\centering
\begin{tabular}{|c|c|c|c|}
\hline
$a/M$ & $\Re\omega_{\text{ana}}/\Re\omega_{\text{num}}$ & $\Im\omega_{\text{ana}}/\Im\omega_{\text{num}}$ \\
\hline
\ 0.9\ \ & \ 1.007\ \ &\ 1.095\ \\
\ 0.95\ \ & \ 1.010\ \ &\ 1.053\ \\
\ 0.99\ \ & \ 1.005\ \ &\ 0.995\ \\
\ 0.995\ \ & \ 1.003\ \ &\ 0.983\ \\
\hline
\end{tabular}
\caption{Massive scalar quasinormal resonances of a near-extremal
Kerr black hole. The data shown is for the mode $l=m=2$ with
$M\mu=0.1$, see also \cite{Dolan}. We display the ratio between the
analytically derived frequencies, $\omega_{\text{ana}}$, and the
numerically computed values, $\omega_{\text{num}}$. The agreement
between the numerical data and the analytical formula (\ref{Eq24})
is quite good already at $a/M=0.9$.} \label{Table2}
\end{table}

\newpage
\section{Summary}

In summary, we have studied analytically the quasinormal mode
spectrum of massive fields in the spacetime of near-extremal
rotating black holes. It was shown that the fundamental resonances
can be expressed in terms of the black-hole physical parameters: the
temperature $T_{BH}$ and the angular velocity $\Omega$. Furthermore,
we have shown that there exists a critical mass parameter,
$\mu_c(l,m)$, below which increasing the mass $\mu$ of the field
increases the oscillation frequency $\Re(\omega)$ of the resonance.
On the other hand, above the critical field mass increasing the mass
$\mu$ of the field increases the damping rate $\Im(\omega)$ of the
mode. We confirmed our analytical results by numerical computations.

\bigskip
\noindent
{\bf ACKNOWLEDGMENTS}
\bigskip

This research is supported by the Meltzer Science Foundation. We
thank Yael Oren, Arbel M. Ongo and Ayelet B. Lata for helpful
discussions.



\begin{thebibliography}{99}

\bibitem{un1} W. Israel, Phys. Rev. {\bf 164}, 1776 (1967); Commun.
Math. Phys. {\bf 8}, 245 (1968).

\bibitem{un2} B. Carter, Phys. Rev. Lett. {\bf 26}, 331 (1971).

\bibitem{un3} S. W. Hawking, Commun. Math. Phys. {\bf 25}, 152 (1972); D. C. Robinson, Phys. Rev. D {\bf 10}, 458 (1974); Phys. Rev. Lett.
{\bf 34}, 905 (1975); J. Isper, Phys. Rev. Lett. {\bf 27}, 529
(1971).

\bibitem{Nollert1} H. P. Nollert, Class. Quantum Grav. {\bf 16}, R159 (1999).

\bibitem{Ber1} E. Berti, V. Cardoso and A. O. Starinets, Class. Quant. Grav. {\bf 26}, 163001
(2009).

\bibitem{Tails1} R.H. Price, Phys. Rev. D {\bf 5}, 2419 (1972);
C. Gundlach, R.H. Price, and J. Pullin, Phys. Rev. D {\bf 49}, 883
(1994); J. Bic\'ak, Gen. Relativ. Gravitation {\bf 3}, 331 (1972).

\bibitem{Tails2} E. S. C. Ching, P. T. Leung, W. M. Suen, and K.
Young, Phys. Rev. Lett. {\bf 74}, 2414 (1995); E. S. C. Ching, P. T.
Leung, W. M. Suen, and K. Young, Phys. Rev. D {\bf 52}, 2118 (1995);
S. Hod and T. Piran, Phys. Rev. D {\bf 58}, 024017 (1998)
[arXiv:gr-qc/9712041]; S. Hod and T. Piran, Phys. Rev. D {\bf 58},
024018 (1998) [arXiv:gr-qc/9801001]; S. Hod and T. Piran, Phys. Rev.
D {\bf 58}, 044018 (1998) [arXiv:gr-qc/9801059]; S. Hod and T.
Piran, Phys. Rev. D {\bf 58}, 024019 (1998) [arXiv:gr-qc/9801060];
S. Hod, Phys. Rev. D {\bf 58}, 104022 (1998) [arXiv:gr-qc/9811032];
S. Hod, Phys. Rev. D {\bf 61}, 024033 (2000) [arXiv:gr-qc/9902072];
S. Hod, Phys. Rev. D {\bf 61}, 064018 (2000) [arXiv:gr-qc/9902073];
L. Barack, Phys. Rev. D {\bf 61}, 024026 (2000); S. Hod, Phys. Rev.
Lett. {\bf 84}, 10 (2000) [arXiv:gr-qc/9907096]; S. Hod, Phys. Rev.
D {\bf 60}, 104053 (1999) [arXiv:gr-qc/9907044]; S. Hod, Class.
Quant. Grav. {\bf 26}, 028001 (2009) [arXiv:0902.0237]; S. Hod,
Class. Quant. Grav. {\bf 18}, 1311 (2001) [arXiv:gr-qc/0008001]; S.
Hod, Phys. Rev. D {\bf 66}, 024001 (2002) [arXiv:gr-qc/0201017]; R.
J. Gleiser, R. H. Price, and J. Pullin, Class. Quant. Grav. {\bf
25}, 072001 (2008); M. Tiglio, L. E. Kidder, and S. A. Teukolsky,
Class. Quant. Grav. {\bf 25}, 105022 (2008); R. Moderski and M.
Rogatko, Phys. Rev. D {\bf 77}, 124007 (2008); X. He and J. Jing,
Nucl. Phys.B {\bf 755}, 313 (2006); H. Koyama and A. Tomimatsu,
Phys. Rev. D {\bf 65}, 084031 (2002); B. Wang, C. Molina, and E.
Abdalla, Phys. Rev. D {\bf 63}, 084001 (2001).

\bibitem{Teuk} S. A. Teukolsky, Phys. Rev. Lett. {\bf 29}, 1114 (1972);
Astrophys. J. {\bf 185}, 635 (1973); W. H. Press and S. A.
Teukolsky, Astrophys. J. {\bf 185}, 649 (1973).

\bibitem{Detw} S. L. Detweiler, in Sources of Gravitational Radiation,
edited by L. Smarr (Cambridge University Press, Cambridge, England,
1979).

\bibitem{HodPRL} S. Hod, Phys. Rev. Lett. 81, 4293 (1998)
[arXiv:gr-qc/9812002].

\bibitem{Gary} G. T. Horowitz and V. E. Hubeny, Phys. Rev. D {\bf 62}, 024027
(2000).

\bibitem{Hod1} S. Hod, Phys. Rev. D {\bf 75}, 064013 (2007) [arXiv:gr-qc/0611004];
S. Hod, Class. and Quant. Grav. {\bf 24}, 4235 (2007)
[arXiv:0705.2306]; A. Gruzinov, arXiv:gr-qc/0705.1725.

\bibitem{Hod2} S. Hod, Phys. Lett. B {\bf 666} 483 (2008) [arXiv:0810.5419];
S. Hod, Phys. Rev. D {\bf 80}, 064004 (2009) [arXiv:0909.0314].

\bibitem{Hod3} S. Hod, Phys. Lett. A {\bf 374}, 2901 (2010) [arXiv:1006.4439].

\bibitem{Will} L. E. Simone and C. M. Will, Class. Quantum Grav. {\bf 9}, 963 (1992).

\bibitem{Siedel} E. Seidel and W. M. Suen, Phys. Rev. Lett. {\bf
66}, 1659 (1991).

\bibitem{Dolan} S. R. Dolan, Phys. Rev. D {\bf 76}, 084001 (2007).

\bibitem{Kerr} R. P. Kerr, Phys. Rev. Lett. {\bf 11}, 237 (1963);
R. H. Boyer and R. W. Lindquist, J. Math. Phys. {\bf 8}, 265 (1967).

\bibitem{Heun} A. Ronveaux, {\it Heun's differential equations}.
(Oxford University Press, Oxford, UK, 1995).

\bibitem{Flam} C. Flammer, {\it Spheroidal Wave Functions} (Stanford
University Press, Stanford, 1957).

\bibitem{Fiz1} P. P. Fiziev, e-print arXiv:0902.1277.

\bibitem{Fiz2} R. S. Borissov and P. P. Fiziev, e-print
arXiv:0903.3617.

\bibitem{Fiz3} P. P. Fiziev, Phys. Rev. D {\bf 80}, 124001 (2009); P. P. Fiziev, Class. Quant. Grav. {\bf 27}, 135001
(2010).

\bibitem{Abram} M. Abramowitz and I. A. Stegun, {\it Handbook of
Mathematical Functions} (Dover Publications, New York, 1970).

\bibitem{Stro} T. Hartman, W. Song, and A. Strominger, JHEP 1003:118 (2010) [arXiv:0908.3909].

\bibitem{Hodcen} S. Hod, Phys. Rev. Lett. {\bf 100}, 121101 (2008)
[arXiv:0805.3873].

\bibitem{DamDer} T. Damour, N. Deruelle and R. Ruffini, Lett. Nuovo Cimento {\bf
15}, 257 (1976).

\bibitem{HodSO} S. Hod and O. Hod, Phys. Rev. D {\bf 81}, 061502 Rapid
communication (2010) [arXiv:0910.0734].

\bibitem{Teukms} S. A. Teukolsky and W. H. Press, Astrophys. J. {\bf 193}, 443 (1974);
A. A. Starobinsky, Zh. Eksp. Teor. Fiz. {\bf 64}, 48 (1973) [Sov.
Phys. JETP {\bf 37}, 28 (1973)]; A. A. Starobinsky and S. M.
Churilov, Zh. Eksp. Teor. Fiz. {\bf 65}, 3 (1973) [Sov. Phys. JETP
{\bf 38}, 1 (1973)]; S. Detweiler, Astrophys. J. {\bf 239}, 292
(1980); S. Hod, Phys. Rev. D {\bf 78}, 084035 (2008)
[arXiv:0811.3806].

\bibitem{Morse} P. M. Morse and H. Feshbach, {\it Methods of
Theoretical Physics} (McGraw-Hill, New York, 1953).

\bibitem{Notedelta} Had we taken $\Re\delta<0$ or $\Im\delta<0$, we would have found
that $\epsilon\gg 1$. In this case a consistent solution of the
resonance condition (\ref{Eq20}) would require
$1/\Gamma(1/2+i\delta-i\varpi+ik)=O(\epsilon^{-1})$. It is
straightforward to show that the black-hole resonances would still
be given by the same analytical formula, see Eq. (\ref{Eq24}) below.

\end{thebibliography}
\end{document}